\documentclass[pra,final,twocolumn]{revtex4-2}
\usepackage{amsmath}
\usepackage{graphicx}
\usepackage{amsfonts}
\usepackage{amssymb}
\usepackage{subfigure}

\begin{document}


\title{On Bell's Dynamical Route to Special Relativity }

\author{Frederick W. Strauch}
\affiliation{ 
Department of Physics, Williams College, Williamstown, MA 01267
}

\date{\today}




\begin{abstract}
This paper develops the approach to special relativity put forward by John S. Bell.  The classical dynamics of an electron orbiting a nucleus in uniform motion is solved analytically and compared to numerical simulations for an accelerated nucleus.  The relativistic phenomena of length contraction and time dilation are shown to result from the electric and magnetic forces on the electron when its motion is analyzed in a single frame of reference.  The relevance of these results for understanding the theory of special relativity is discussed.
\end{abstract}

\maketitle



\section{\label{sec:Introduction}Introduction}

In 1976, John Stewart Bell presented his thoughts on how to teach special relativity \cite{[{}][{. Reprinted in }]Bell1976, *Bell2004}.  He proposed a numerical calculation, performed in a single frame of reference, to demonstrate length contraction and time dilation.  This calculation would track the classical orbit of an electron moving about a nucleus that is accelerated from rest to some velocity $v$.  Provided the acceleration is sufficiently gentle, an initially circular orbit of period $T_0$ will contract to an ellipse with axes in ratio of $1:\sqrt{1-v^2/c^2}$ and a final period of $T=T_0/\sqrt{1-v^2/c^2}$, where $c$ is the speed of light.  The orbital contraction is illustrated in Fig. 1.  Further analysis of the orbit would motivate the Lorentz transformations of special relativity.

This dynamical approach contrasts with the kinematical approach of Einstein that is traditionally used to introduce special relativity.  Bell's approach was inspired by the investigations of the ``aether theorists'': FitzGerald, Larmor, Lorentz, and Poincar{\'e}.  The primary virtue of this approach is that it illustrates (albeit in a classical context) how the relativistic effects of length contraction and time dilation can be explained not by the effects of the aether or due to the geometry of space-time, but as consequences of Maxwell's equations applied to moving atoms.  

The purpose of this paper is to provide a detailed pedagogical treatment of Bell's orbital contraction problem, suitable for a broad range of audiences.  The results developed here could be used in courses in modern physics as well as advanced courses in mechanics or electromagnetism.  These could be used to supplement the traditional approach to introducing special relativity.  As Bell wrote \cite{Bell1976},
\begin{quote}
There is no intention here to make any reservation whatever about the power and precision of Einstein's approach.  But in my opinion there is also something to be said for taking students along the road made by FitzGerald, Larmor, Lorentz, and Poincar{\'e}.  The longer road sometimes gives more familiarity with the country.
\end{quote}
This path has become increasingly relevant to philosophers of physics and may be of interest to all students and teachers of relativity.

\begin{figure}
\includegraphics[width = 2in]{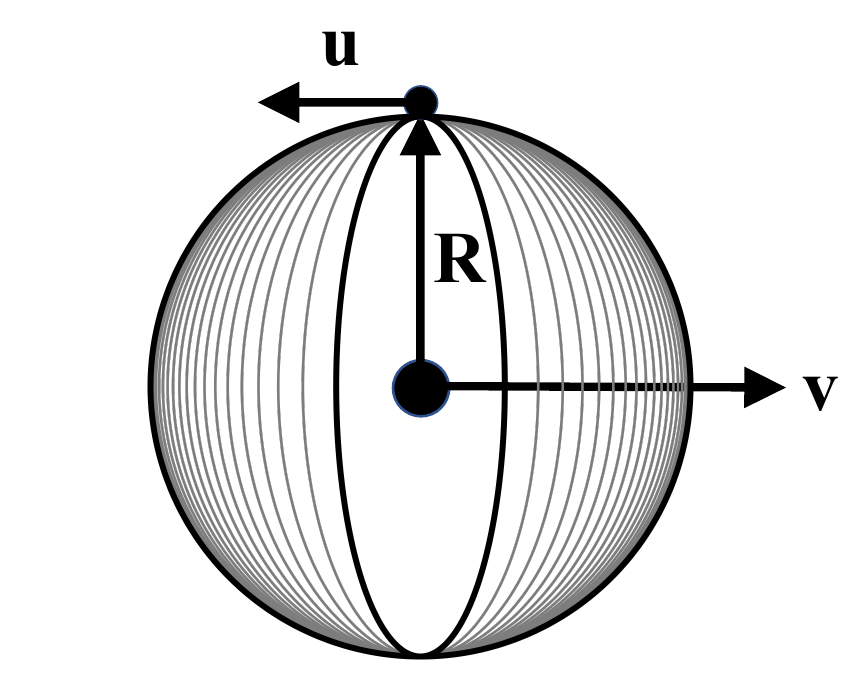}
\caption{\label{fig1} Bell's orbital contraction problem.  An electron with position ${\bf R}$ and velocity ${\bf u}$ orbits a nucleus moving with velocity ${\bf v}$.  An initially circular orbit, for a nucleus with velocity ${\bf v = 0}$, will contract along the direction of motion when the nucleus is gently accelerated to some final velocity.  The final orbit is an ellipse with axes in ratio $1:\sqrt{1-v^2/c^2}$. }
\end{figure}

This paper is organized as follows.  In Sec.~II, I provide a short discussion of the history and the motivation for Bell's orbital contraction problem.  The main results of this study are found in Secs.~III and IV.  Here the equations of motion for the electron's orbit are solved analytically (approximately and exactly).  For completeness, the results of numerical solutions for Bell's original proposal are presented in Sec.~V.  In Sec.~VI, I conclude by discussing various points of interest---pedagogical and philosophical---that appear along this route to special relativity.  Finally, certain pedagogical notes and technical details of the numerical method are presented in the Appendices.   



\section{Background}

The traditional approach to teaching special relativity begins by discussing the 1887 experiment of Michelson and Morley and its failure to observe the motion of the Earth with respect to the ``luminiferous aether''  \cite{tipler2012modern,*taylor2015modern}.  It is then asserted that the correct explanation for this null experiment was given by Einstein's analysis in 1905.  Some texts will mention the Lorentz-FitzGerald hypothesis, namely that an interaction between matter and the aether serves to physically contract an object along its direction of motion.  This hypothesis, which is more accurately understood as a ``deformation hypothesis'' \cite{brown2001origins}, is usually characterized as an \emph{ad hoc} attempt to preserve the aether, an absolute frame of reference whose rejection required an Einstein-level of genius, either by the actual Einstein or perhaps the French mathematician Henri Poincar{\'e} \cite{french2017special}.  

This story, while true in many respects, has several flaws.  First, it does not do justice to the achievements of the Dutch physicist Hendrik Lorentz and the perspicacity of the Irish physicist George FitzGerald.  Second, it fails to provide a full context for Einstein's 1905 paper.  Finally, it misses altogether the contributions of FitzGerald's countryman and friend Joseph Larmor.

It is not hard to find more complete stories, such as that found in Chapter 1 of the famous text on relativity by Wolfgang Pauli \cite{pauli2013theory}, the scientific biography of Einstein by Abraham Pais \cite{Pais1986}, or the text by Harvey Brown \cite{brown2005}.  A long-unknown piece of this story was revealed by the discovery \cite{hunt88} that FitzGerald's hypothesis of 1889 was inspired by a 1888 calculation by Oliver Heaviside \cite{[{}][{.  \\ This paper and many other original works from the late 1800s and early 1900s can be found via www.wikisource.org}]Heaviside1888}.  Here I provide a short summary of the historical background relevant to Bell's approach (see also Ref. \cite{Bell92}). 

Heaviside's result, which can be derived from Maxwell's equations alone \cite{jefimenko1994direct}, states that the magnitude of the electric field of a charge $q$ in uniform motion is reduced in its direction of motion and increased in the direction perpendicular to its motion.  This result is usually found in texts on electricity and magnetism (e.g. \cite{griffiths_2017}) but not in introductory texts on special relativity (with some some notable exceptions \cite{french2017special}). For a point ${\bf r}$ at time $t$, with a displacement ${\bf R} = {\bf r} - {\bf v } t$ from the charge moving with velocity ${\bf v}$, the electric potential is given by
\begin{equation}
    V({\bf r}, t) = \frac{q}{4\pi \epsilon_0} \frac{1}{\sqrt{1 - v^2 \sin^2 \theta/c^2}} \frac{1}{R},
    \label{Vheaviside}
\end{equation}
where $\theta$ is the angle between ${\bf R}$ and ${\bf v}$.  The electric field is given by
\begin{equation}
    {\bf E}({\bf r},t) = \frac{q}{4\pi \epsilon_0} \frac{1-v^2/c^2}{(1-v^2 \sin^2 \theta / c^2)^{3/2}} \frac{\bf{R}}{R^3}.
    \label{heaviside}
\end{equation}

FitzGerald based his hypothesis on this result, writing \cite{fitz1889ether}
\begin{quote}
    I would suggest that almost the only hypothesis that can [explain the Michelson-Morley experiment] is that the length of material bodies changes, according as they are moving through the ether or across it, by an amount depending on the square of the ratio of their velocity to that of light.  We know that electric forces are affected by the motion of the electrified bodies relative to the ether, and it seems a not improbable supposition that the molecular forces are affected by the motion, and that the size of a body alters consequently.
\end{quote}
That is, if matter were so constituted that its dimensions were due to electromagnetic forces, matter in motion could be subject to some change in dimension. This ``not improbable supposition'' was made more plausible by subsequent work by Searle \cite{searle1896xvii, *searle1897xlii} and Morton \cite{morton1896liii}, who showed that the equilibrium conditions for a moving conductor involve the equipotential surfaces for Eq.~(\ref{Vheaviside}), namely the surfaces for which
\begin{equation}
R \sqrt{1 - v^2 \sin^2 \theta/c^2} = \frac{1}{\gamma} \left[ \gamma^2 (x-v t)^2+ y^2 + z^2\right]^{1/2} 
\end{equation}
is constant, where $\gamma = 1/\sqrt{1-v^2/c^2}$ is the usual relativistic factor.  A short history of this ``Heaviside ellipsoid'' can be found in \cite{red2004image}.

The deformation hypothesis was put forward independently by Lorentz in 1892 \cite{Relativity_Papers}, who went on to develop his famous transformations over the next ten years, along with contemporaneous work by Larmor \cite{Larmor1900}.  After discussing the use of the Lorentz transformations  \footnote{The modern form of the Lorentz transformations were not fully justified until the 1905 works of Einstein and Poincar{\'e}; see Ch. 6 of Pais (Ref. 6) or Ch. 4 of Brown (Ref. 7).  Thus, the arguments given by Lorentz and Larmor for length contraction were incomplete.} to connect electromagnetic systems in relative motion, Larmor made the following remarkable comment in his {\it Aether and Matter} (1900):
\begin{quote}
    let us consider ... a pair of electrons of opposite signs describing steady circular orbits round each other in a position of rest ... when this pair is moving through the aether with velocity $v$ in a direction lying in the plane of their orbits, these orbits relative to the translatory motion will be flattened along the direction of $v$ to ellipticity $1-{\frac{1}{2}}v^{2}/c^{2}$, while there will be a first-order retardation of phase in each orbital motion when the electron is in front of the mean position combined with acceleration when behind it so that on the whole the period will be changed only in the second-order $1+{\frac {1}{2}}v^{2}/c^{2}$. 
\end{quote}

Bell proposed to turn Larmor's observation into a dynamical calculation of the orbital contraction.  This requires the Lorentz force law along with one additional piece, namely the change in the electron's inertial mass when it is in motion.  Historically, this was proposed by Lorentz in 1904 \cite{Relativity_Papers} and further analyzed by Poincar{\'e} in 1905 \cite{Miller73}.  Lorentz's proposal was based on (at least \cite{[{}][{.  Holton identifies eleven distinct hypotheses.}]holton60} two further hypotheses: first, that there is intrinsic structure to the electron (that is itself subject to length contraction) and second, that the entire mass of the electron is of electromagnetic origin.  The modern understanding of the relativistic force law, based on Einstein's 1905 analysis \cite{Relativity_Papers}, was established by Planck in 1906 \cite{[{}][{.  See also Sec. 29 of Pauli (Ref. 5).}]max1906prinzip}.

The orbital contraction problem considered in this paper is as follows.  Consider a classical model of the atom consisting of a nucleus with $q = + Z e$ in uniform motion with velocity ${\bf v}$.   An electron of charge $-e$ with (rest) mass $m$ and velocity ${\bf u} = d {\bf r}/dt$ will be subject to the equation of motion
\begin{equation}
    \frac{d}{dt} \left( \frac{m {\bf u}}{\sqrt{1-u^2/c^2}} \right) = -e \left( {\bf E} + {\bf u} \times {\bf B} \right),
    \label{forcelaw}
\end{equation}
where the electric field ${\bf E}$ of the nucleus is given by Eq.~(\ref{heaviside}) and the corresponding magnetic field is
\begin{equation}
    {\bf B}({\bf r}, t) = \frac{1}{c^2} {\bf v} \times {\bf E}({\bf r}, t).
    \label{heavisideb}
\end{equation}
Analytical solutions to the electron's equation of motion will be found, and we will see that the circular orbits that are observed when $v=0$ deform into length-contracted and time-dilated elliptical orbits when the atom is in motion.  Bell actually proposed a more involved numerical calculation using the electric and magnetic fields for an accelerating nucleus.  Should the nucleus undergo an ``adiabatic'' acceleration, the electron orbit will dynamically contract from a circle to an ellipse, as illustrated in Fig.~1.  Numerical simulations of adiabatic and non-adiabatic accelerations will be shown in Sec. V.

The essential point of Bell's approach is that one can learn the core lessons of special relativity--the Lorentz transformations, length contraction, and time dilation---by a dynamical calculation performed in a single frame of reference.  By contrast, the traditional approach explains these effects by reference to inertial coordinate systems, using either the kinematic properties of space and time (as developed by Einstein in 1905) or the corresponding four-dimensional geometry of spacetime (as given by Minkowski in 1908) \cite{Relativity_Papers}.  That both approaches agree is a necessary consequence of Lorentz invariance.

There are a few important observations to be made about this approach, which Bell called the ``Lorentzian pedagogy'' \cite{[{A critical discussion of the ``Lorentzian pedagogy'' described by Bell can be found in }][{ and }]brown2001origin,*[{}][{.  See also Brown (Ref. 7).}]brown2003michelson}.  First, the aether plays no significant role in this calculation (although it was seen as essential for Lorentz and most other pre-Einstein theorists); the only ingredients are Maxwell's equations and the (relativistic) force law of Eq.~(\ref{forcelaw}), expressed in a single frame of reference.  As Lorentz and Poincar{\'e} obtained this force law using an outdated electromagnetic view of matter, Bell argued that this force law could be taken as a fundamental result from experiment.  A recent argument \cite{walstad2018relativistic}, however, shows that the form of the force law, specifically the expression of the relativistic momentum, can be found using a novel modification of Einstein's 1906 argument for $E=m c^2$ \cite{einstein1906prinzip}.

Second, Bell proposed to introduce the Lorentz transformations after calculating the orbit of the electron, whereas Larmor and Lorentz freely used these transformations in their investigations.  The calculation presented here will use the relativistic force law of Eq.~(\ref{forcelaw}) but will not make direct use of the Lorentz transformations.  The electron orbits for the case of uniform motion will be found analytically in Sec.~IV.  Given the analytical form of these orbits, it will be shown that the deformed orbits for the moving atom are precisely mapped to the circular orbit for the atom at rest by the Lorentz transformations.

Finally, the model of matter proposed by Bell is, of course, inadequate, insofar as it involves a classical model of the atom, an externally imposed trajectory for the nucleus, and the neglect of any radiation effects.  Nevertheless, the orbital contraction problem can play an important role in understanding the conceptual structure of special relativity.  On this point, Einstein, writing in his {\em Autobiographical Notes} \cite{Einstein1949}, observed that the theory involves
\begin{quotation}
\noindent two kinds of physical things, i.e., (1) measuring rods and clocks, (2) all other things, e.g., the electro-magnetic field, the material point, etc.  This, in a certain sense, is inconsistent; strictly speaking measuring rods and clocks would have to be represented as solutions of the basic equations (objects consisting of moving atomic configurations), not, as it were, as theoretically self-sufficient entities.
\end{quotation}
The solutions obtained here address this point directly; further aspects of the ``Lorentzian pedagogy''  will be discussed in Sec.~VI.  

\section{Approximate Orbit for Nucleus in Uniform Motion}

I begin by solving the orbital contraction problem for slowly moving orbits, using techniques accessible to students of modern physics.  I first note that the equation of motion Eq.~(\ref{forcelaw}) can written as $d {\bf p}/dt = {\bf F}$ where
\begin{equation}
    {\bf p} = \frac{ m {\bf u}}{\sqrt{1-u^2/c^2}}
    \label{momentum}
\end{equation}
is the relativistic momentum.  The slow orbit approximation is $| {\bf{u}} - {\bf{v}}| \ll c$.  Restricting to orbits in the $x-y$ plane and setting $x = \xi + v t$ so that $u_x = v + d \xi /dt$, one can Taylor expand Eq.~(\ref{momentum}) to lowest order in $d\xi/dt$ and $dy/dt$.  Taking the time derivative results in the approximations
\begin{align}
\frac{d p_x}{dt} &\approx m (1-v^2/c^2)^{-3/2} \frac{d^2 \xi}{dt^2}, \nonumber \\
\frac{d p_y}{dt} &\approx m (1-v^2/c^2)^{-1/2} \frac{d^2 y}{dt^2}.
\label{dpdt}
\end{align}

The corresponding components of the Lorentz force ${\bf F} =  -e \left( {\bf E} + {\bf u} \times {\bf B} \right)$ can be expressed as
\begin{align}
F_x &= - e \left(E_x + \frac{v}{c^2} u_y E_y \right) \approx - e E_x, \nonumber \\
F_y &= - e \left(E_y - \frac{v}{c^2} u_x E_y \right) \approx - e (1 - v^2/c^2) E_y,
\label{fapprox}
\end{align}
where I have used the fact that $B_z = v E_y/c^2$ as well as the slow orbit approximation.  The electric field given by Eq.~(\ref{heaviside}) depends on the combination 
\begin{equation}
\tilde{r} = \sqrt{ \gamma^2 (x-v t)^2 + y^2 } = \sqrt{ \gamma^2 \xi^2 + y^2}.
\end{equation}
Circular orbits, when $v=0$, result from the constraint $\tilde{r} = r_0$, where $r_0$ is a constant.  Remarkably, this constraint can be used for $v \ne 0$.  Doing so yields the electric field components 
\begin{equation}
E_x = \gamma (E_0 /r_0) \xi \quad \mbox{and} \quad E_y = \gamma (E_0 /r_0) y ,
\label{Eapprox}
\end{equation}
where I have defined $E_0 = Z e / (4 \pi \epsilon_0 r_0^2)$.

The equations of motion for slow orbits are found by substituting Eq.~(\ref{Eapprox}) into Eq.~(\ref{fapprox}) and equating with Eqs. (\ref{dpdt}):
\begin{align}
m \gamma^3 \frac{d^2 \xi}{dt^2} &= - e \gamma (E_0 / r_0)  \xi, \nonumber \\
m \gamma \frac{d^2 y}{dt^2} &=- e \gamma^{-1} (E_0 /r_0) y.
\end{align}
These are the equations for simple harmonic motion.  Defining the angular frequency $\omega_0$ (appropriate for slow circular orbits with $v=0$)
\begin{equation}
\omega_0^2 = \frac{e}{m} \frac{E_0}{r_0} = \frac{1}{4 \pi \epsilon_0} \frac{Z e^2}{m r_0^3},
\label{freq0}
\end{equation}
the equations can be simplified to 
\begin{align}
\frac{d^2 \xi}{dt^2} &= - \omega_0^2 (1 - v^2/c^2) \xi, \nonumber \\
\frac{d^2 y}{dt^2} &= - \omega_0^2 (1- v^2/c^2) y.
\end{align}

To satisfy the constraint $\tilde{r} = r_0$, the solutions must be of the form
\begin{eqnarray}
x(t) &=& v t + \gamma^{-1} r_0 \cos (\omega t), \nonumber \\
y(t) &=& r_0 \sin (\omega t),
\label{circleorbit}
\end{eqnarray}
where $\omega = \omega_0 \sqrt{1 - v^2/c^2}$.  When $v = 0$, this is a circular orbit.  For $v \ne 0$ this is the length-contracted, time-dilated elliptical orbit.  

The length contraction of the orbit results from the constraint $\tilde{r} = r_0$, which arises from the form of the electric field in Eq.~(\ref{heaviside}).  The time dilation of the orbit, however, arises in two different ways.  For the $x$-component, it arises from the approximation for $dp_x/dt$ in Eq.~(\ref{dpdt})  (in particular via the relativistic ``longitudinal mass,'' using archaic terminology).  For the $y$-component, however, it arises from the magnetic field's contribution to $F_y$ in Eq.~(\ref{fapprox}).  Taken together, this dynamical calculation points toward an electromagnetic origin of these relativistic effects.

\section{Exact Orbits for Nucleus in Uniform Motion}

While the previous section solved the orbital contraction problem in the limit that the orbital speed is small, it is possible to calculate an analytic solution for arbitrary orbital velocities.  I shall begin with the relativistic motion of an electron about a nucleus at rest, and then proceed to the more general case of uniform motion.  The connection of these two cases by the Lorentz transformations will then be discussed.  This method of calculation, appropriate for advanced undergraduates, serves to highlight the power of the Lorentz transformations for a somewhat involved problem in relativistic classical dynamics. (Some suggestions toward that end are provided in Appendix A).

\subsection{Circular Orbit about Nucleus at Rest}

The circular orbits for the case when $v=0$ can be solved relatively easily.  Here, both the radius $r = |{\bf r}| = r_0$ and the velocity $u = |{\bf u}| = r_0 \omega$ are constant in magnitude, while $ {\bf u} \cdot d {\bf u}/dt = 0$, and $d {\bf u}/dt = - \omega^2 {\bf r}$.  The equation of motion Eq.~(\ref{forcelaw}) reduces to 
\begin{equation}
\frac{m r_0 \omega^2}{\sqrt{1 - r_0^2 \omega^2/c^2}}  = e E_0.
\label{eom_circle}
\end{equation}
Using the expression for $\omega_0$ in Eq.~(\ref{freq0}), defining the dimensionless quantity $\eta = r_0 \omega_0 /c$, and squaring Eq.~(\ref{eom_circle}), leads to the following equation for $\omega$:
\begin{equation}
\left( \frac{\omega}{\omega_0} \right)^4  + \eta^2 \left( \frac{\omega}{\omega_0}\right)^2  - 1 = 0.
\label{freq_rest}
\end{equation}
The appropriate square root is given by
\begin{equation}
\omega_1 = \omega_0 \left( \sqrt{1+\eta^4/4} - \eta^2/2 \right)^{1/2}.  
\label{freq_circle}
\end{equation}
Note that the slow orbit approximation used in the previous section is justified when $\eta \ll 1$.

A few other observations about $\omega_1$ are in order.  The parameter $\eta$ can be expressed as
\begin{equation}
\eta^2 =  \frac{r_0^2 \omega_0^2}{c^2} = \frac{1}{4\pi \epsilon_0} \frac{Z e^2}{r_0} \frac{1}{m c^2}  = \frac{Z r_e}{r_0},
\end{equation}
where $r_e$ is known as the classical electron radius, and is defined by
\begin{equation}
r_e = \frac{1}{4\pi \epsilon_0} \frac{e^2}{m c^2}.
\end{equation}
In the limit of a large orbital radius $r_0 \gg Z r_e$, $\eta \to 0$ and $\omega_1 \to \omega_0 \propto r_0^{-3/2}$.  One can also consider the opposite limit of a small orbital radius $r_0 \ll Z r_e$ (in which $\omega_0$ and $\eta$ increase as inverse powers of $r_0$):
\begin{equation}
\omega_1 \to \omega_0 / \eta = \frac{c}{r_0}.
\end{equation}
As expected, the orbital velocity $u = r_0 \omega_1$ is always less than the speed of light.

\begin{figure*}
\includegraphics[width = 6.5in]{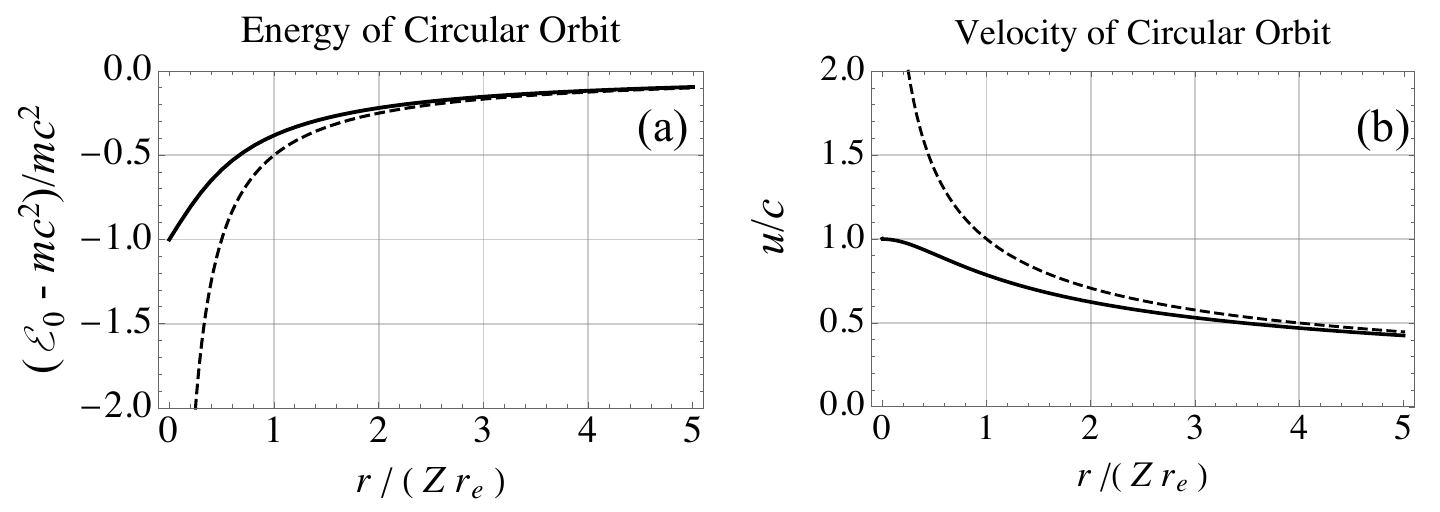}
\caption{\label{fig2} (a) The energy $\mathcal{E}_0 - m c^2$ and (b) the velocity $u = r \omega_1$ of the circular orbit as a function of the orbital radius $r/(Z r_e)$, where $r_e$ is the electron radius, $Z$ is the atomic number of the nucleus, and the energy and velocity are in units of $m c^2$ and $c$, respectively.  Also shown (dashed) are the non-relativistic results for the energy $- Z e^2 / (8 \pi \epsilon_0 r)$ and the orbital velocity $\sqrt{Z e^2 /(4 \pi \epsilon_0 m r)}$. }
\end{figure*}

As discussed in the Appendix, the equation of motion Eq.~(\ref{forcelaw}), when $v=0$, has the conserved energy
\begin{equation}
\mathcal{E}(v=0) = m c^2 \left( \frac{1}{\sqrt{1-u^2/c^2}} \right) -  \frac{1}{4 \pi \epsilon_0}  \frac{Z e^2}{r}
\label{energy_circle}
\end{equation}
For circular orbits (with $r=r_0$ and $u = r_0 \omega_1$), this expression for the energy can be simplified using the definitions for $\omega_0$ and $\omega_1$.  This results in
\begin{equation}
\mathcal{E}(v=0) \to \mathcal{E}_0 = m c^2 \left( \sqrt{1 + \eta^4/4}  - \frac{1}{2} \eta^2 \right).
\label{energy_orbit}
\end{equation}
For slow orbits, $\eta \ll 1$ and thus 
\begin{equation}
\mathcal{E}_0 \approx m c^2 (1 - \frac{1}{2} \eta^2) = m c^2 - \frac{1}{8 \pi \epsilon_0} \frac{Z e^2}{r_0},
\end{equation}
while for fast orbits, $\eta \gg 1$ and 
\begin{equation}
\mathcal{E}_0 \approx m c^2 \frac{1}{\eta^2} = 4 \pi \epsilon_0 (m c^2)^2 \frac{r_0}{Z e^2}.
\end{equation}
The results for the orbital velocity and the energy are shown in Fig.~\ref{fig2}.

\subsection{Elliptical Orbit about Nucleus in Motion}

The results of Sec.~III suggest that, when the nucleus is in motion, the electron's orbit can be written as
\begin{eqnarray}
x(t) &=& v t + \gamma^{-1} r_0 \cos \theta(t), \nonumber \\
y(t) &=& r_0 \sin \theta(t),
\label{ellipse}
\end{eqnarray}
where $\theta(t)$ is a function of $t$ to be determined.  To proceed, one could substitute Eq.~(\ref{ellipse}) into Eq.~(\ref{forcelaw}) to obtain a second-order differential equation for $\theta(t)$.  It is advantageous, however, to identify a constant of motion like the energy, and thus find a first-order differential equation. 

As shown in the Appendix, the equation of motion Eq.~(\ref{forcelaw}), when $v \ne 0$, has the constant of motion 
\begin{equation}
\mathcal{E}(v) = \gamma \left( \frac{m c^2 - m v u_x}{\sqrt{1-u^2/c^2}} \right) - \frac{1}{4 \pi \epsilon_0} \frac{Z e^2}{\tilde{r}}.
\label{constant}
\end{equation}
For the elliptical orbits with $\tilde{r} = r_0$, the constant of motion can be simplified to
\begin{equation}
K = \frac{1 - u_x v / c^2} {\sqrt{1-u^2/c^2}} = \mbox{constant}.
\label{constant2}
 \end{equation}
This can then be used to find $\theta(t)$.

Taking the time derivative of Eq.~(\ref{ellipse}) yields
\begin{eqnarray}
 u_x &=& v - \gamma^{-1} r_0 \sin \theta \ \dot{\theta} \nonumber \\
 u_y &=& r_0 \cos \theta \ \dot{\theta}.
 \end{eqnarray}
Substituting $u_x$ and $u_y$ into Eq.~(\ref{constant2}), one can show (after some algebra) that 
\begin{equation}
K = \frac{1}{\gamma} \frac{ 1 + \tau \ \dot{\theta} \sin \theta }{ \left[ \left(1 + \tau \ \dot{\theta} \sin \theta \right)^2 - r_0^2 \gamma^2 \dot{\theta}^2/c^2 \right]^{1/2}},
\end{equation}
where $\tau = \gamma r_0 v/c^2$. Evaluating $K$ at a time $t_0$ such that $\theta(t=t_0) = 0$ leads to the first-order differential equation
\begin{equation}
\dot{\theta} = \frac{\dot{\theta}_0}{1 - \tau \ \dot{\theta}_0 \sin \theta}, 
\end{equation}
where $\dot{\theta}_0 = \dot{\theta}(t_0)$.
As a separable equation, this has the implicit solution
\begin{equation}
\theta(t) + \tau \ \dot{\theta_0} \cos \theta(t) = \dot{\theta}_0 t,
\label{kepler}
\end{equation}
from which we find $t_0 = \tau$.  Remarkably, this expression for $\theta(t)$ is a form of Kepler's equation for the elliptical orbits in a gravitational potential \cite{Orlando18}. 

It remains to find the frequency of the orbit, and to relate it to that of the electron orbiting the nucleus at rest.  First, by setting $\theta(T)-\theta(0) = 2 \pi$, where $ T = 2 \pi / \omega$ for some frequency $\omega$, we see that $\dot{\theta}_0 = \omega$. Second, one can show (as discussed in the Appendix, and after some more algebra) that
\begin{equation}
    \frac{m c^2 \tau \omega^2}{\sqrt{1-r_0^2 \gamma^2 \omega^2/c^2}} = e E_0 v/\gamma.
    \label{eom_ellipse}
\end{equation}
Using the expressions for $\omega_0$ and $\eta$, and squaring Eq.~(\ref{eom_ellipse}) leads to the following equation for $\omega$:
\begin{equation}
\gamma^4 \left( \frac{\omega}{\omega_0}\right)^4 + \eta^2 \gamma^2 \left( \frac{\omega}{\omega_0} \right)^2 - 1 = 0.
\end{equation}
Note that this equation is equivalent to Eq.~(\ref{freq_rest}) under the replacement $\omega_0 \to \omega_0 / \gamma$.  Thus, the appropriate square root is given by
\begin{eqnarray}
\omega_2 &=& \omega_0 \sqrt{1-v^2/c^2} \left( \sqrt{1+\eta^4/4} - \eta^2/2 \right)^{1/2} \nonumber \\
&=& \omega_1 \sqrt{1-v^2/c^2}.
\label{freqeq}
\end{eqnarray}

In summary, the equations of motion for an electron orbiting a nucleus in uniform motion can be solved exactly, yielding the elliptical orbit of Eq.~(\ref{ellipse}), with $\theta(t)$ given by (\ref{kepler}), and $\omega_2 = \dot{\theta}_0$ given by (\ref{freqeq}).  This orbit is contracted along the direction of motion by the factor $\sqrt{1-v^2/c^2}$, while the frequency of the orbit is reduced by the same factor.   

\subsection{Role of the Lorentz Transformations}

So far, the orbital contraction problem has been solved without any use of the Lorentz transformations.  Using these, however, one can show that the elliptical orbit for the moving atom transforms to the circular orbit for the atom at rest.  That is, applying the Lorentz transformations to the elliptical orbit Eq.~(\ref{ellipse}) from the frame in which the nucleus is in motion ($x,y,t$) to the frame in which the nucleus is at rest ($x',y',t'$) yields
\begin{eqnarray}
x' &=& \gamma (x - v t) = r_0 \cos \theta, \nonumber \\
y' &=& y = r_0 \sin \theta, \nonumber \\
t' &=& \gamma (t - v x/c^2) = \gamma^{-1} (t - \tau \cos \theta).
\end{eqnarray}
This last equation can be simplified using Eq.~(\ref{kepler}): 
\begin{equation}
    t' = \gamma^{-1} \frac{\theta}{\omega_2} = \frac{\theta}{\omega_1}.
\end{equation}
Thus, $\theta = \omega_1 t'$, and we recover the circular orbit for $ x'(t'), y'(t')$, in agreement with Sec. IV.A.  

An observant reader might ask, what compels us to match our initial circular orbit, with $r = r_0$, to a contracted elliptical orbit, with $\tilde{r} = r_0$, with the same value of $r_0$?  That is, one might worry that the argument given so far only justifies a {\it deformation} of the orbit, but not the precise contraction given by the Lorentz transformations.  To answer that question, consider the constants of motion given by Eqs.~(\ref{energy_circle}) and (\ref{constant}).  Evaluating these for $r=\tilde{r}=r_0$, one obtains
\begin{equation}
\mathcal{E}(v) = \mathcal{E}(0) = \mathcal{E}_0,
\end{equation}
for arbitrary values of $v$. The equality of these energy functions is precisely what one expects for an ``adiabatic'' acceleration of the nucleus.  That this energy remains approximately constant for a specific trajectory for the nucleus will be demonstrated in the following section. 

The equality of the energy functions can be further understood by using the Lorentz transformations for energy and momentum.  In the frame in which the nucleus is at rest, the energy of the circular orbit can be written as $\mathcal{E}(0) = E' - e V'$, where $V' = Z e /(4 \pi \epsilon_0 r')$ is the scalar potential and $E' = m c^2 / \sqrt{1 - u'^2/c^2}$.  Using the appropriate Lorentz transformation \cite{griffiths_2017} to express this energy in the frame in which the nucleus is in motion yields
\begin{equation}
 \mathcal{E}(0) = \gamma (E - v p_x) - e \gamma (V - v A_x),
\end{equation}
\\
where $E = m c^2 / \sqrt{1 -  u^2/c^2}$, $V$ is the potential from Eq.~(\ref{Vheaviside}) (with $q = Z e$) and $A_x = v V/c^2$ is the vector potential.  This expression is in complete agreement with the form of $\mathcal{E}(v)$ in Eq.~(\ref{constant}).

In summary, I have calculated the orbit of the electron subject to Eq.~(\ref{forcelaw}) for a moving classical atom using the constant of motion of Eq.~(\ref{constant}).  This constant of motion corresponds to the energy of the electron.  This energy will remain constant for an adiabatic acceleration of the atom, for which an initially circular orbit for an electron contracts into an elliptical orbit, as illustrated in Fig.~1.  The corresponding orbits are precisely related by the Lorentz transformations of special relativity.

\bigskip

\section{Numerical Orbits for Accelerated Nucleus}

The actual problem proposed by Bell was to numerically calculate the motion for an accelerating nucleus.  This requires the full electric and magnetic fields \cite{griffiths_2017}
\begin{equation}
    {\bf E}({\bf r}, t) = \frac{q}{4\pi\epsilon_0}\frac{R}{({\bf R} \cdot {\bf U})^3} \left[ (c^2 - v^2) {\bf U} + {\bf R} \times ({\bf U} \times {\bf a}) \right]
\end{equation}
and
\begin{equation}
    {\bf B}({\bf r},t) = \frac{1}{c} \hat{{\bf R}} \times {\bf E}({\bf r},t).
\end{equation}
Here ${\bf R} = {\bf r} - {\bf r}_n(t_r)$, where ${\bf r}_n(t)$ is the position of the nucleus, ${\bf U} = c \hat{{\bf R}} - {\bf v}$, $t_r = t - R/c$ is the retarded time, and the velocity ${\bf v} = d {\bf r}_n/dt$ and the acceleration ${\bf a} = d^2 {\bf r}_n/dt^2$ are evaluated at time $t_r$.  

\begin{figure*}
\includegraphics[width = 6.5in]{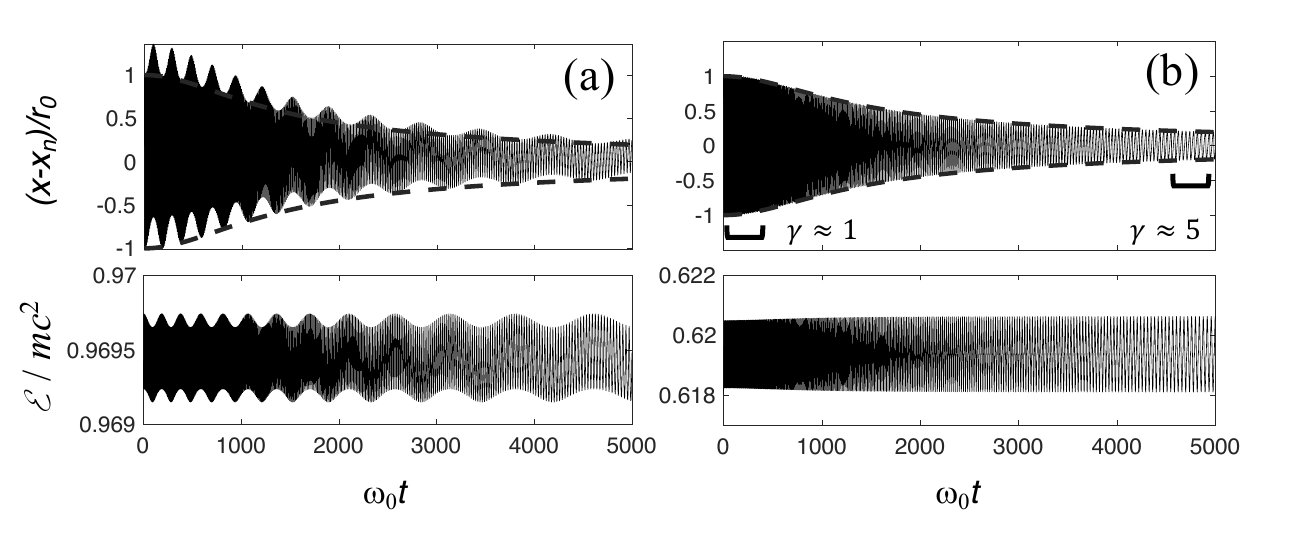}
\caption{\label{fig3} Relative coordinate $x-x_n$ (top) and instantaneous energy $\mathcal{E}[v(t)]$ (bottom) of the electron as a function of time $t$, in units of $r_0$, $m c^2$,  and $1/\omega_0$, respectively, for an accelerated nucleus with (a) $\eta = 0.25$ and $x_0/r_0 = 4000$ and (b) $\eta = 1$ and $x_0/r_0 = 1000$.  The values of $\eta$ and $x_0/r_0$ were chosen to provide the same velocity $v(t)/c = \omega_0 t / \sqrt{(\omega_0 t)^2 + (x_0 \eta/r_0)^2}$ for the nucleus.   Also shown in the top panels is the contraction factor $1/\gamma = \sqrt{1-v(t)^2/c^2}$ (dashed).  The motion begins with $\gamma \approx 1$  (with $v/c \approx 0$) and ends with $\gamma \approx 5$  (with $v/c \approx 0.98$). The values of $\mathcal{E}_0/m c^2$ from Eq.~(\ref{energy_orbit}) corresponding to (a) and (b) are $0.9692$ and $0.618$, respectively.  The modulation seen in (a) is due to the non-adiabatic acceleration of the nucleus for the weakly bound electron orbit with $\eta = 0.25$.  Conversely, the absence of such modulation in (b) is due to the adiabatic acceleration of the nucleus for the strongly bound electron orbit $\eta = 1$. }	
\end{figure*}

I let the trajectory of the nucleus be given by ${\bf r}_n = (x_n,0)$, where
\begin{equation}
    x_n(t) = \left\{ \begin{array}{ll}
    x_0 & \mbox{for} \ t < 0, \\
    \sqrt{x_0^2 + c^2 t^2} & \mbox{for} \ t > 0,
    \end{array} \right.
\end{equation}
This has been termed ``truncated hyperbolic motion'', and the corresponding electric field has recently been investigated \cite{franklin2015electric}.  The retarded time $t_r$ is the following explicit function of $t$ and ${\bf r}$:
\begin{widetext}
\begin{equation}
    t_r = \left\{ \begin{array}{ll}
     \frac{1}{2} \frac{(c^2 t^2 - r^2 - x_0^2)}{(c^2 t^2 - x^2)} t +\frac{1}{2} \frac{x/c}{(c^2 t^2 - x^2)} \left[(c^2 t^2-r^2-x_0)^2 + 4 x_0^2 (c^2 t^2 - x^2) \right]^{1/2} & \mbox{for} \ t > \sqrt{(x-x_0)^2 + y^2}/c,  \\
t - \sqrt{(x-x_0)^2 + y^2}/c &  \mbox{for} \  t < \sqrt{(x-x_0)^2 + y^2}/c.
\end{array} \right.
\end{equation}
\end{widetext}
Note that the acceleration for hyperbolic motion is initially $c^2/x_0$.

Using the length scale $r_0$ and the time scale $1/\omega_0$, the motion is fully determined by the dimensionless parameters $\eta$ and $x_0/r_0$. The numerical method is described in Appendix B, and was verified by calculating the exact orbits (and constants of motion) for constant velocity found in Sec.~IV; see supplemental material for computer code \footnote{The {\tt MATLAB} code is available at [url to be inserted by AIPP]}.  Numerically integrating the time-dependent equations of motion produces the results for $x(t) - x_n(t)$ shown in the top panels of Fig.~\ref{fig3}(a) and (b), for two different sets of $x_0/r_0$ and $\eta$.  As the nucleus accelerates, the electron's orbit contracts along the direction of motion and the frequency decreases, in agreement with the results of the previous section.  The contraction and time dilation of the orbit follow the instantaneous relativistic factor $\gamma(t) = 1/\sqrt{1-v(t)^2/c^2}$.  However, the trajectory in Fig.~\ref{fig3}(a), with $\eta = 0.25$ exhibits a modulation on top of the orbital oscillations.  This modulation is diminished for $\eta=1$, as seen in Fig.~\ref{fig3}(b).  

The same modulation is seen in the instantaneous energy $\mathcal{E}[(v(t)]$, as shown in the bottom panels of Fig.~\ref{fig3}(a) and (b).  Note that the energies remain close to the values of $\mathcal{E}_0$ predicted by Eq.~(\ref{energy_orbit}).  In fact, the diminished modulation for $\eta=1$ can be understood by considering these values.  As discussed in Sec. IV, small values of $\eta$ correspond to large radii with weakly bound orbits, while larger values of $\eta$ correspond to smaller radii with more strongly bound orbits.  Thus, the more strongly bound orbit can handle higher accelerations.  Note that if the length scale $r_0 = a_0/Z$, where $a_0$ is the Bohr radius and $Z$ the atomic number of the nucleus, then $\eta$ can be written as $Z \alpha$, where $\alpha \approx 1/137$ is the fine structure constant.  Calculation of the orbital contraction for a hydrogen atom would require much smaller accelerations of the nucleus, in agreement with Bell's discussion.

\begin{figure*}
\includegraphics[width = 6.5in]{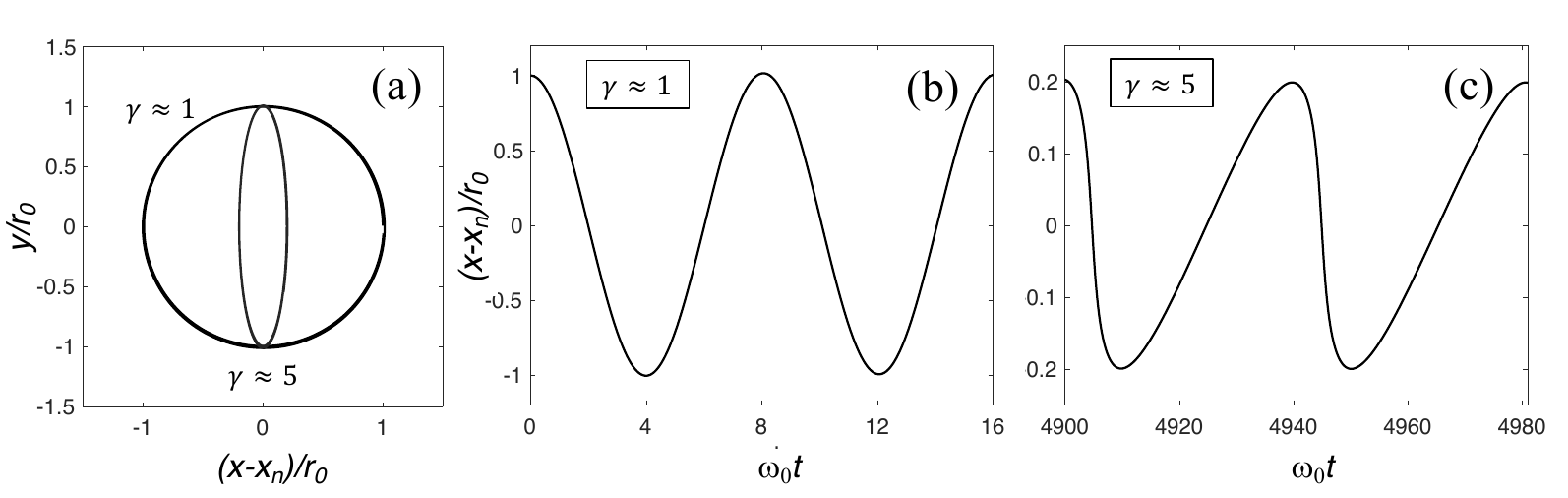}
\caption{\label{fig4} (a) Electron orbits about the accelerated nucleus, calculated for $\eta = 1$ and $x_0/r_0 = 1000$ for relativistic factors $\gamma \approx 1$ (with $v/c \approx 0$) and $\gamma \approx 5$ (with $v/c \approx 0.98$).  The corresponding relative coordinate $x-x_n$ of the electron as a function of time $t$, in units of $r_0$ and $1/\omega_0$, for (a) $\gamma \approx 1$ and (b) $\gamma \approx 5$.}
\end{figure*}

Two representative orbits of the trajectory for $\eta = 1$, with relativistic factors $\gamma \approx 1$ and $\gamma \approx 5$, are shown in Fig.~\ref{fig4}(a).   The corresponding results for $x(t)-x_n(t)$ are shown in Fig.~\ref{fig4}(b) and (c).  Using the results of Sec. IV, for $\gamma \approx 1$ and $\eta=1$ the period is $2\pi/\omega_1 \approx 8 / \omega_0$, while for $\gamma \approx 5$ the period is $2 \pi / \omega_2 \approx 40 / \omega_0$, in agreement with the numerical results shown in in Fig.~\ref{fig4}(b) and Fig.~\ref{fig4}(c).

\section{Discussion}

We have followed Bell's path, or at least this particular journey along it, to its conclusion.  Proceeding further, one can encounter recent developments in the study of the history and philosophy of physics.  In particular, the contrast between Bell's ``Lorentzian'' and the traditional approaches has been analyzed by Brown and co-workers \cite{brown2001origin, brown2001origins,brown2003michelson, brown2005, brown2006minkowski}.  He has challenged the tradition (on a number of points), prompting a lively debate on our understanding of theories of spacetime \cite{janssen2009drawing,pooley2013substantivalist,brown2021dynamical}.  One of the central topics in this debate is the nature of explanations \cite{dieks2009bottom}.  The ``Lorentzian'' path takes a ``constructive theory'' (or ``bottom-up'') approach that grounds explanations in physical or dynamical terms, whereas the traditional story follows a ``principle theory'' (or ``top-down'') approach proceeds from kinematical or geometrical considerations \cite{[{The characterization of ``principle'' and ``constructive'' theories was given by Einstein in 1919, as discussed in }][{. The terms ``top-down'' and 'bottom-up'' are due to Dieks (in Ref. 34).  See also Sec. 5.2 of Brown (Ref. 7): }]howard2019einstein}.  Bell encourages us to better understand the relationship between (or at least recognize the validity and utility of) these two approaches \cite{Bell92}.  

From this perspective, there are several pedagogical advantages to supplementing the traditional approach using the treatment developed here.  A primary advantage is that it provides a dynamical account of the relativistic effects that ``moving objects shrink'' and ``moving clocks run slow'' (albeit using a classical model of the atom).  This account can be seen as a consistency check on the traditional approach, to ensure that rods and clocks, as traditionally used in special relativity, actually do what they are kinematically supposed to do \cite{[{As noted in Sec. II, Einstein understood that it was desirable to describe physical rods and clocks using a complete dynamical theory.  Note that no classical account can fully achieve this goal, but also that the Lorentz-invariance of the full dynamical theory will ensure consistency.  For a historical analysis of Einstein's thinking on this matter, see }][{. See also Sec. 4 of Brown and Pooley (Ref. 23), Secs. 7.1 and 8.4 of Brown (Ref. 7)}. ]giovanelli2014but}.  Similar calculations for simple models of rods and clocks ``built'' from configurations of point charges have been considered previously \cite{jefimenko96b, *Redzic2015a, *Redzic2015b}; constructive \cite{miller2010} and qualitative  \cite{Nelson2015} arguments have also been put forward.

Dynamical approaches are subject to the criticism that they depend on Lorentz invariance through the relativistic force law.  That this law is, in fact, the correct force law requires its own justification.  Historically, this law was obtained by Planck using Einstein's ``principle theory'' of 1905, in which Lorentz invariance follows from the relativity and light postulates.  The contemporaneous account for such a force law given by the ``constructive theory'' of Lorentz and Poincar{\'e} required a number of additional assumptions \cite{janssen2009drawing}, in addition to their commitment to an unobservable aether \footnote{Both Lorentz and Poincar{\'e} saw the aether as the medium for electromagnetic phenomena.  This role was replaced by spacetime in Einstein's theory.  See Pais (Ref. 6) and Secs. 4.7 and 8.4 of Brown (Ref. 7)}.  As noted in Sec. II, however, the dynamical approach can be strengthened on this point.  Specifically, the argument given in \cite{walstad2018relativistic} (based on an argument originally due to Poincar{\'e} and Einstein) shows that the form of the relativistic momentum can be deduced from the requirement that the center-of-mass of a closed system at rest remains constant.  

Another advantage of this approach is that it demonstrates how relativistic effects can be understood within a single frame of reference.  The power of Lorentz invariance can then be demonstrated by showing how the elliptical orbit and the Kepler equation Eq.~(\ref{kepler}) can be derived via the Lorentz transformations as an {\em active} transformation from $(x',t'$) to ($x,t$) of a circular orbit about a nucleus at rest.  

A final advantage is that it provides some historical context, illustrating how relativistic effects had been appreciated (if only partially) before 1905 and illuminating (if only partly) how Lorentz and Poincar{\'e} could remain attached to the aether after 1905.  From our perspective, their attachment to such an unnecessary hypothesis may seem peculiar.  From their perspective, it was Einstein's elevation of unproven hypotheses to fundamental postulates that seemed peculiar 
\cite{[{This was certainly true for Lorentz; the ultimate opinion of Poincar{\'e} is less clear.  See Sec. II of }][{ and Sec. II of }]goldberg1969lorentz,*goldberg1967henri,*[{ For an interesting perspective, see the ``imaginary discussion'' in }][{}]darrigol1996electrodynamic}. 

In conclusion, this route can offer students a glimpse of the historical and philosophical context behind special relativity.  As observed by Brown and Read \cite{brown2021dynamical}:
\begin{quote}
We expect undergraduates to imbibe in their first course on relativity theory a profound insight largely obscure to {\em all} the nineteenth century giants, including Maxwell, Lorentz, Larmor, and Poincar{\'e}: the physical meaning of inertial coordinate transformations.  It was Einstein in 1905 who was the first to understand the physics of such transformations, and the  fact that they are neither {\it a priori} nor conventional.
\end{quote}
It is hoped that the treatment developed here will help future students better understand the physical content of special relativity.

\begin{acknowledgments}

I gratefully acknowledge many fruitful discussions with Keith McPartland and Bill Wootters on various aspects of special relativity.  I also thank the three anonymous referees for their comments on an earlier version of this manuscript.

\end{acknowledgments}

\appendix

\bigskip

\section{Pedagogical Notes}

Bell's orbital contraction problem could be used in a number of different ways.  It is this author's opinion that the orbital contraction problem could be used in a modern physics course after introducing relativistic dynamics.  While there is an argument in favor of a fully constructive approach over the traditional approach to the teaching of special relativity, including elements of both seems the most prudent.  Further discussion can be found in Miller \cite{miller2010}, who argues that it is worthwhile to use
\begin{quotation}
\noindent a dynamical constructive approach even if one is going to argue in hindsight that the concepts are best seen to flow directly from the principles or from the geometry of Minkowski spacetime. However, it is probably not suitable to replace the usual principle-based approach with a constructive approach in most courses on special relativity because of the need to rely on a fairly sophisticated understanding of electromagnetism and the need to develop the relativistic version of Newton’s second law before being able to derive time dilation constructively. A better option from a pedagogical point of view may be to refer to the constructive approach in qualitative terms as an alternative to the principle-based approach and motivate students to follow up the details if they are interested. The understanding of a subject is usually enhanced if it is presented from different points of view.
\end{quotation}

For students of modern physics, the qualitative discussion in Sec.~II might suffice.  Interested students could also read Bell's 1976 paper \cite{Bell1976} and his tribute to FitzGerald \cite{Bell92}.  A quantitative approach would start by solving the relativistic form of Newton's second law in one dimension for a constant force (which yields the hyperbolic motion used in Sec.~VI).  One could then study the two-dimensional motion for a charged particle in a constant magnetic field.  Finally, given the Heaviside result for the electric and magnetic fields, the approximate calculation of Sec.~III could be performed.  

Advanced students could perform the calculations of Sec.~IV.  Specifically, they should verify the constant of motions $\mathcal{E}(0)$, given by Eq~(\ref{energy_circle}), and $\mathcal{E}(v)$, given by Eq.~(\ref{constant}).   This can be done for the former (where $v=0$) by taking the dot product of the equation of motion Eq.~(\ref{forcelaw}) 
 \begin{equation}
{\bf u} \cdot \frac{ d {\bf p}}{dt} = - e {\bf u} \cdot {\bf E},
\end{equation}
and first showing that that left-hand-side can be expressed as
\begin{eqnarray}
{\bf u} \cdot \frac{ d {\bf p}}{dt}  &=& \frac{m}{(1-u^2/c^2)^{3/2}} \left( {\bf u} \cdot \frac{d {\bf u}}{dt} \right) \nonumber \\
&=& m c^2 \frac{d}{dt} \left( \frac{1}{\sqrt{1-u^2/c^2}} \right).
\label{udpdt}
\end{eqnarray}
Then, the right-hand-side can be  
\begin{eqnarray}
- e {\bf u} \cdot {\bf E} &=&
- \frac{Z e^2}{4 \pi \epsilon_0} \frac{1}{r^3} \left( {\bf r} \cdot \frac{d {\bf r}}{dt}  \right) \nonumber \\
&=& \frac{d}{dt} \left( \frac{1}{4 \pi \epsilon_0} \frac{Z e^2}{r} \right).
\end{eqnarray}
Combining these two results demonstrates that $d\mathcal{E}(0)/dt = 0$.

To verify that $\mathcal{E}(v)$ is a constant of motion, one can combine 
\begin{equation}
\frac{d}{dt} \left( \frac{m c^2}{\sqrt{1-u^2/c^2}} \right) = - e {\bf E} \cdot {\bf u}.
\label{eqnrg}
\end{equation}
with the results
\begin{equation}
\frac{d}{dt} \left( \frac{m u_x} {\sqrt{1-u^2/c^2}} \right) = - e (E_x + \frac{v u_y}{c^2} E_y )
\end{equation}
and
\begin{equation}
\frac{d}{dt} \left( \frac{1}{4\pi \epsilon_0} \frac{Ze^2}{\tilde{r}} \right) = - e \gamma (u_x - v) E_x - e \gamma^{-1} u_y E_y
\end{equation}
to show that $d \mathcal{E}(v)/dt = 0$.  

Finally, to verify that the elliptical orbit satisfies the equation of motion, one substitutes Eqs.~(\ref{ellipse}) and (\ref{kepler}) into Eq.~(\ref{eqnrg}) to obtain Eq.~(\ref{eom_ellipse}).  

\bigskip

\section{Numerical Method}

The numerical results presented in Sec. VI  use a second-order split-step integration method \cite{mclachlan_quispel_2002}.  

Given the electron's position $(x,y)$ and momentum $(p_x,p_y)$ at time $t$, the position and momentum at time $t+\Delta t$ are found by the following sequence of updates:

\begin{enumerate}
\item{The position is updated by 
\begin{eqnarray}
x^* &=& x  + (\Delta t/2) p_x /  \sqrt{m^2 + (p_x^2 + p_y^2)/c^2}, \nonumber \\
y^* &=& y  + (\Delta t/2) p_y /  \sqrt{m^2 + (p_x^2 + p_y^2)/c^2}.
\end{eqnarray}}
\item{The electric and magnetic fields are then evaluated at the position $(x^*,y^*)$ and the time $t^* = t + \Delta t/2$. }
\item{The momentum is updated according to the electric field by
\begin{eqnarray}
p_x &=& p_x - e E_x (\Delta t/2), \nonumber \\
p_y &=& p_y - e E_y (\Delta t/2).
\end{eqnarray}}
\item{The momentum is updated according to the magnetic field by the rotation
\begin{eqnarray}
p_x^* &=& p_x \cos \varphi  - p_y \sin \varphi, \nonumber \\
p_y^* &=& p_y \cos \varphi + p_x \sin \varphi,
\end{eqnarray}
where 
\begin{equation}
    \varphi = e B_z \Delta t / \sqrt{m^2 + (p_x^2 + p_y^2)/c^2}.
\end{equation}}
\item{The (final) momentum is obtained by the electric field update
\begin{eqnarray}
p_x &=& p_x ^*- e E_x (\Delta t/2), \nonumber \\
p_y &=& p_y ^*- e E_y (\Delta t/2).
\end{eqnarray}}
\item{The (final) position is obtained by the update
\begin{eqnarray}
x &=& x^*  + (\Delta t/2) p_x /  \sqrt{m^2 + (p_x^2 + p_y^2)/c^2}, \nonumber \\
y &=& y^*  + (\Delta t/2) p_y /  \sqrt{m^2 + (p_x^2 + p_y^2)/c^2}.
\end{eqnarray}}
\end{enumerate}

The initial conditions used for the simulations shown in the text are
\begin{eqnarray}
    x(t=0) &=& x_0 + r_0, \nonumber \\
    y(t=0) &=& 0, \nonumber \\
    p_x(t=0) &=& 0, \nonumber \\
    p_y(t=0) &=& m r_0 \omega_1 / \sqrt{1-r_0^2 \omega_1^2/c^2},
\end{eqnarray}
where $\omega_1$ is given by Eq.~(\ref{freq_circle}).  The actual code uses the dimensionless variables $x/r_0$, $y/r_0$, $p_x/(m r_0 \omega_0)$, $p_y / (m r_0 \omega_0)$, and $\omega_0 t$. 

\section{Author Declarations}
The authors have no conflicts to disclose.

\bibliography{bell_paper}

\end{document}